\documentclass[graybox]{svmult}

\usepackage{mathptmx}       % selects Times Roman as basic font
\usepackage{helvet}         % selects Helvetica as sans-serif font
\usepackage{courier}        % selects Courier as typewriter font
\usepackage{type1cm}        % activate if the above 3 fonts are
\usepackage{makeidx}         % allows index generation
\usepackage{graphicx}        % standard LaTeX graphics tool
\usepackage{multicol}        % used for the two-column index
\usepackage[bottom]{footmisc}% places footnotes at page bottom
\usepackage[misc]{ifsym}
\usepackage{algorithmicx}

\makeindex             % used for the subject index
\begin{document}

\title*{Mining the Social Media Data for a Bottom-Up Evaluation of Walkability}
\author{Christian Berzi, Andrea Gorrini, Giuseppe Vizzari}
\authorrunning{Berzi, C., Gorrini, A., Vizzari, G.} 
\institute{Christian Berzi \and Andrea Gorrini \and Giuseppe Vizzari (\Letter) \at CSAI-Complex Systems and Artificial Intelligence research centre\\Department of Informatics, Systems and Communications, University of Milano-Bicocca\\Viale Sarca 336 - Building U14, 20126 Milan (ITALY)\\\email{{name.surname}@disco.unimib.it} }

\maketitle

\abstract*{}

\abstract{Urbanization represents a huge opportunity for computer applications enabling cities to be managed more efficiently while, at the same time, improving the life quality of their citizens. One of the potential application of this kind of systems is a bottom-up evaluation of the level of walkability of the city (namely the level of usefulness, comfort, safety and attractiveness of an urban area for walking). This is based on the usage of data from social media for the computation of structured indicators describing the actual usage of areas by pedestrians. This paper will present an experimentation of analysis of data about the city of Milano (Italy) acquired from Flickr and Foursquare. The over 500 thousand points, which represent the photos and the POIs collected from the above mentioned social meda, were clustered through an iterative approach based on the DBSCAN algorithm, in order to achieve homogeneous areas defined by the actual activity of inhabitants and tourists rather than by a top down administrative procedure and to supply useful indications on the level of walkability of the city of Milan.}

\section{Introduction}
\label{sec:1}

\emph{Urbanisation} is one of the most significant trend of the 21\textsuperscript{st} Century. Today over half of the world's population lives in urban areas and all the regions of the world are expected to urbanise further over the coming decades: by 2050 66\% of the world's population will reside in cities \cite{nations2014world}. This demographic trend requires Institutions to more effectively design and plan the cities and to improve the quality of life of the inhabitants, considering the rising demand of services and infrastructures. One of the most essential element of new development agendas for future cities is based on the need to adopt sustainable mobility strategies, based for example on increasing public transport services in order to decongest vehicular traffic volumes and reduce the level of environmental pollution. Facing this trend, advanced urban planning activities \cite{wefering2013guidelines} are shifting toward a focus on pedestrian mobility and walkability \cite{abley2005walkability}, namely how friendly the urban environment is for walking. 

In general, the term walkability \cite{Ewing2009} refers to those infrastructural-design elements which guarantee the comfort, safety and accessibility of the urban settings for walking, considering also the specific needs of those people with limited mobility. More in detail, Jeff Speck \cite{Speck2012} has recently proposed a \textit{General Theory of Walkability}, which defines the four essential criteria for the assessment of the walkability level of urban environments:

\begin{itemize}
\item \textit{Usefulness}: the urban territory should be designed and planned in order to guarantee the presence of diverse public services, commercial activities and green areas within a walkable distance of 150 meters;
\item \textit{Comfort}: side-walks should be designed according to a series of standard criteria of quality and accessibility (e.g., adequate width of the pavement, ramps for people with impaired mobility, legible road sign for way-finding);
\item \textit{Safety}: streets should be designed in order to guarantee the safety of pedestrians (e.g., bumpers in proximity of the zebra, sufficient illumination systems, adequate horizontal and vertical signage);
\item \textit{Attractiveness}: the city should be designed to have a policentric structure, with several and distinctive areas of attraction for both the citizens and the visitors; this is based on the presence of points of interest and events, the quality of the architectural streetscape and the vitality of the social context.
\end{itemize} 

According to the multi-faceted definition of walkability depicted by Speck, the assessment of the pedestrian friendliness of a determined urban area should comprise the evaluation of: (\textit{i}) \textit{structured indicators} related to the topographical and spatial elements of the area (e.g., presence of public services, quality of road infrastructures) and to the census of the socio-demographic characteristics of its inhabitants; (\textit{ii}) \textit{behavioural indicators} related to how the spatial features of the area influences the actual behaviours of people while walking (e.g., problematic level of services, pedestrian exposure to risks); (\textit{iii}) \textit{subjective indicators} focused on the evaluation of the city users about the level of walkability of the area (e.g., perceived comfort and safety). Within a multi-disciplinary approach (e.g., urban studies, architecture, urban sociology, environmental psychology, computer science), a wide range of methods and techniques have been developed and applied to empirically measure the level of walkability of the urban environments: field observations \cite{Day2005}; questionnaires \cite{checklist}; GIS-analyses \cite{Agampatian2014}; web-based applications \cite{duncan2013s} (see Tab. \ref{tab1} for a schematic summary).

\begin{table}[]
\centering
\caption{A schematic summary of the methods developed for walkability assessment, according to different criteria and indicators.}
\label{tab1}
\begin{tabular}{p{4cm}p{4cm}p{3.5cm}}
\hline
\textbf{Methods} 												       & \textbf{Assessment Criteria} & \textbf{Assessment Indicators} \\ \hline \hline
\textit{Field Observations} \cite{Day2005}  						   & Comfort and safety           & Behavioural 				   \\ \hline
\textit{Questionnaires}   \cite{checklist}                             & Comfort and safety           & Subjective                     \\ \hline
\textit{GIS-analyses} \cite{Agampatian2014}               		       & Usefulness 			      & Structured                     \\ \hline
\textit{Web Applications} \cite{duncan2013s}					       & Comfort and Safety           & Subjective                     \\ \hline
\textit{Social Media Data} \cite{Quercia2014,Quercia2015}     	   & Comfort and Attractiveness       & Structured and  Subjective                    \\ \hline
\end{tabular}
\end{table}

In this framework, the most novel and promising approach for walkability assessment is the one porposed by \textit{Urban Informatics} \cite{foth2011urban}, a relatively new term invented to define a category of computer-based applications somehow gravitating around urban studies, and incorporating different types of knowledge and skills ranging from computer science, architecture and  social science. Urban Informatics approach is based  on the recent developments of advanced information and communication technologies (ICTs) and on the possibility to gather and analyse large scale heterogeneous data (e.g., institutional, user generated social media data). This scenario enables the use of geo-referred social media data for a bottom-up evalution of walkability, directly based on user's experience of the urban environment (without requiring ad-hoc infrastructures, surveys or observations). In particular, the contributions proposed by Daniele Quercia \cite{Quercia2014,Quercia2015} provide a characterization of the urban areas in terms of comfort and pleasantness for walking path recommendation employing user generated contents. 

In this vein, we propose a method to automatically identify and characterize the attractive entities within the city of Milan (Italy), by using some data gathered from Flickr and Foursquare. The rational of the approach is based on two pillars: (\textit{i}) the usage of social media data for identifying and characterising with a bottom-up  perspective particular areas within the city (not unlike in \cite{brindley2017generating}) considering the popularity factor (also associated to different times of day, and days of week, potentially correlated to special events) motivating the choice of taking a picture and uploading it on Flickr; (\textit{ii}) the possibility of interpreting this kind of proxy for attractiveness of the area within the above overall walkability assessment framework. The bottom-up nature of this approach provides results that represent an interesting integration to existing, top down, and more traditional techniques for walkability assessment and already available institutional data and previous urban plans.

The enabling data which set the current work are presented in Section \ref{sec:2}. Section \ref{sec:3} and \ref{sec:4} respectively present the identification and the characterization processes of the identified entities of the city. The paper concludes with remarks about the achieved results and their future use for a bottom-up evaluation of walkability through social media data. 

\section{Enabling Data}
\label{sec:2}

Thanks to the increasing popularity of social media web sites and the parallel diffusion of smart devices embedded with camera, high speed Internet connection and GPS, massive social media sources of geo-referenced data about user behaviours have recently become available. This work considers two particular social media sources, that are Flickr and Foursquare. A preliminary analysis revealed that within a reasonable bounding box including the city of Milano and relevant surroundings (e.g. including Milano Fiera - Rho, adjacent to the area that hosted Expo 2015 and regularly hosting additional fairs) about 450 thousand Flickr photos, and related metadata, are present. Additionally, approximately 50 thousand Foursquare points of interest (POI) are also present in the relevant area. The metadata derived from the two datasets are complementary: in fact Flickr tags are a free form, bottom up type of description of the photo content (and potentially also describing the place/context in which the photo was taken) whereas Foursquare POIs are associated to a centrally defined taxonomy, representing a more authoritative form of description. The main motivation behind this choice is that a potentially interesting urban areas/events should attract people that take photos and upload them on a social media like Flickr or they should contain points of interest (e.g. gardens, monuments, events) accessible on a social media like Foursquare.

\section{City entities identification} 
\label{sec:3}

We chose to use the term \textit{entity} instead of (urban) area of interest\cite{HU2015240} since it may refer to any type of element within the city such as a square, an avenue, a building, a monument, but also an event (associated to a place, but a distinct concept). The approach employed to identify the entities within the city, analogously as the above cited\cite{HU2015240}, is based on the application of a clustering technique on the above mentioned social media data. The proposed characterization of the identified entities is based on a set of features (e.g. temporal features, textual features) extracted from the elements (social media data) that are associated to those identified entities.

DBSCAN (Density-Based Spatial Clustering of Applications with Noise \cite{ester1996density}) is a density-based clustering algorithm which is suited to this specific situation (e.g. good efficiency on large databases, often employed for analysis of spatial data, discovery of clusters with arbitrary shape, no need to specify a priori the number of clusters to identify). Therefore, this algorithm was chosen to identify the entities within the city: it allows to identify and delineate regions of high density (i.e. city areas where there are many photos or POIs) that are separated by regions of lower density. The intuition is that around areas of interest there are regions separating them from potentially nearby other areas, so a sort of “border” can be identified around the area. 

\begin{figure}[th]
\begin{center}
\includegraphics[width=\textwidth]{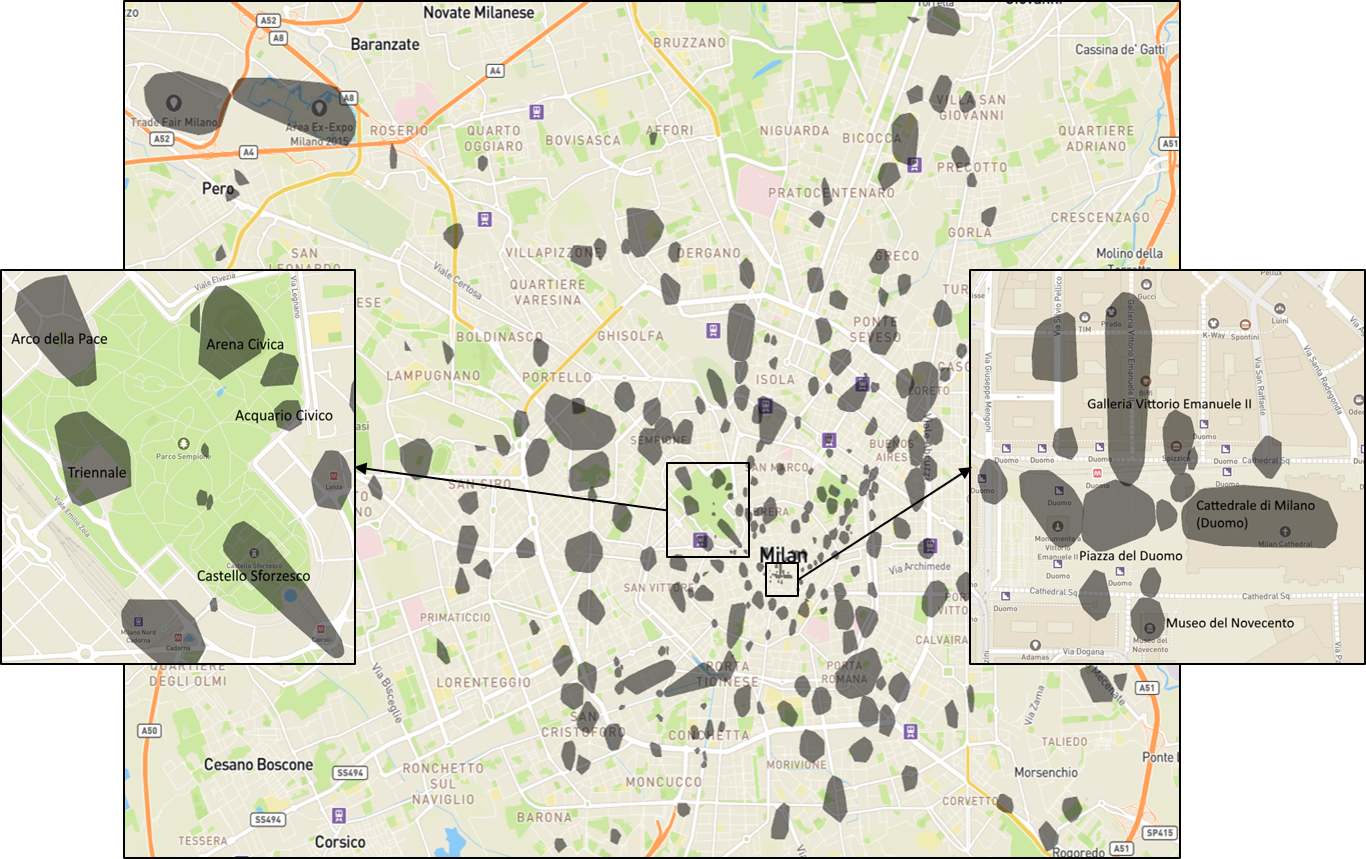}
\caption{The clusters identified within the whole city (notice the Rho Fiera area in the North Western part of the map) and two examples of clusters in the city center (i.e. the Sforzesco Castle - on the left, and the Cathedral area - on the right).}
\label{fig:1}
\end{center}
\end{figure}

The heterogeneous density that characterizes Milano toward the center of the city does not allow to apply a “one-shot” algorithm that covers the entire city. Indeed, a first run of DBSCAN allows to achieve a satisfactory result outside the city center, by identifying some peripheral entities such as Expo 2015, Milano Fiera, San Siro Stadium, Milano Bovisa and so on. On the other hand, the whole center of Milano is represented by a single cluster (i.e. a single entity), that can be considered an unsatisfactory result. Of course, it is possible to tune the sensitivity of DBSCAN according to the density of the center of Milano but, in this case, the algorithm would produce the opposite situation, by achieving an unsatisfactory result outside the center of the city and a satisfactory result within the center of the city. Consequently, with the purpose to produce a satisfactory result in the whole city, the proposed solution is to employ an iterative approach, by re-applying DBSCAN on the clusters for which an unsatisfactory result is achieved by the previous iteration, adjusting each time the sensitivity of the algorithm according to the density of the currently considered area. At the end of the entire process, each identified cluster refers to a potentially interesting entity within the city of Milano (see Fig. \ref{fig:1}).

A quantitative evaluation approach has been proposed to identify potentially interesting entities within the city of Milano. A validity index (i.e. silhouette coefficient) is calculated to measure the quality of the clustering process through the different iterations that are conducted. The silhouette coefficient is a composite index that reflects both the compactness (i.e. how closely related are objects in a cluster) and the separation (i.e. how distinct or well separated a cluster is from other clusters) of clusters (i.e. the entities within the city) \cite{han2011data}.

\begin{figure}[th]
\begin{center}
\includegraphics[width=\textwidth]{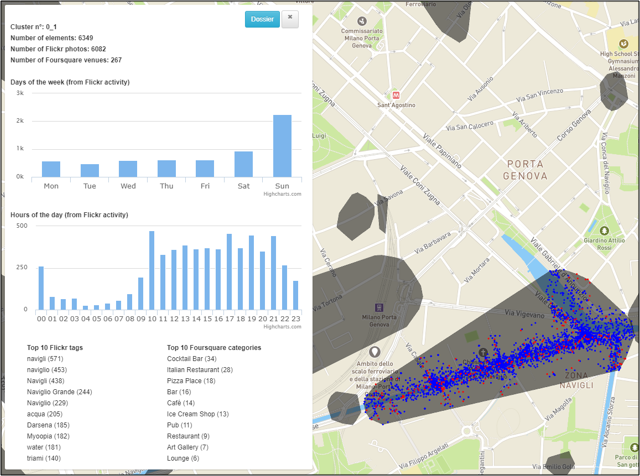}
\caption{The detailed description of a single cluster (i.e. the Navigli area), with indication on the day of week and the time of day in which Flickr photos were taken, and the top ten Flickr tags and Foursquare POI categories.}
\label{fig:2}
\end{center}
\end{figure}

\section{City Entities Characterization}
\label{sec:4}

The identified entities in the city have been characterized with a set of features extracted from the elements (Flickr photos and Foursquare POIs) that are contained within those entities (see Fig. \ref{fig:2}, as an example of the characterisation process of a single cluster - the Navigli area). In particular, for each identified entity, the following features are computed:

\begin{itemize}
\item Flickr temporal features: they are extracted by using the timestamp of when photos are taken. In particular, the following two temporal features are considered: (\textit{i}) number of photos taken during the different days of the week (e.g. Monday, Tuesday, Wednesday); (\textit{ii}) number of photos taken during the different hours of the day, 24h format.
\item Flickr photo tags: the most frequent tags that are used to tag the photos on Flickr. With the purpose to consider the most significant tags, a stop-words-removing phase is included to remove the tags that are not considered significant, such as the language stop words (i.e. function words) and the considered domain stop words. The second one refers to a manually created list of words which must not be considered, since they do not refer to the city entities but they are automatically added by some Flickr extensions of other social media, for instance tags such as: $\mathit{instagram\_id}=\left\langle \mathit{ID}\right\rangle, \mathit{Foursquare\_id}=\left\langle \mathit{ID}\right\rangle, \mathit{Facebook\_id}=\left\langle \mathit{ID}\right\rangle, \mathit{Flickr\_mobile}, \mathit{geotagged}$. \\Furthermore, an algebraic model for representing unstructured data (Vector Space Model \cite{salton1975vector}) is used since it allows to calculate how important a tag is for a specific entity of the city, by using the tf-idf (i.e. term frequency inverse document frequency) weighting.	
\item Foursquare POI categories: the number of points of interest which are associated to each of the 10 main Foursquare categories, that are \textit{Arts \& Entertainment, College \& University, Event, Food, Night-life Spot, Outdoors \& Recreation, Professional \& Other Places, Residence, Shop \& Service, Travel \& Transport}.
\item The number of photos that are taken over time (i.e. in the course of the years), in relation with the most frequent tags. Figure 2 shows a graphic representation of this feature for the cluster associated with the \textit{Navigli area}.
\end{itemize}

\section{Conclusions and Future Works}
\label{sec:5}

Within the Urban Informatics approach, the present work is based on the analysis of social media data (Flickr and Foursquare data) for a bottom-up assessment of the level of walkability of the city of Milano. This is focused on studying the attractiveness of the city for inhabitants and visitors (thanks to the presence of points of interest, the organization of events and the vitality of the social context) to evaluate its overall pedestrian friendliness. The bottom-up nature of this approach provides results that represent an interesting integration to existing, top down, and more traditional techniques for walkability assessment and already available institutional data and previous urban plans (e.g., institutional data about the presence of public services within a walkable distance of 150 meters, the topographical features of the spatial setting, the localisation of pedestrian-car accidents). The proposed multi-layer approach for walkability assessment could be then translated into practice by designers, who would create people-centred optimal environments.

An interesting future work might be the implementation of the labelling process of the identified entities within the city. Through a manual inspection of most of the identified entities, it is apparent that many entities can be labelled simply by using the most frequent tag used in that entity or, in some cases, a tag within the first 10 most frequent tags. On the other hand, it is not always the case that the most frequent tag is the best one and, moreover, some of the identified entities are related to events that frequently take place in an area of the city rather than actual places so, in this cases, tags would not really be descriptive of the environment. Thus, the most frequent tags list can certainly be a good starting point for the labelling task, even though this task requires a more structured process rather than simply use the most frequent tag as the label of the entity. For instance, this process might include some NLP (Natural Language Processing) techniques (e.g. entity linking) and some external linguistic resources (e.g. thesaurus, ontologies) to support the choice of a particular tag (or a combination of tags) as the label of the entity.
 
%\bibliographystyle{spmpsci}
%\bibliography{tgfbibs_tgf}

%alternatively, you can input the references manually. Keep to the format described in 
%Mathematical and Physical Sciences.pdf
%\input{referenc}
\end{document}